# The pyrochlore Ho$_2$Ti$_2$O$_7$: Synthesis, crystal growth and stoichiometry

A. Ghasemi[*], A. Scheie, J. Kindervater, S.M. Koohpayeh

*Institute for Quantum Matter, Department of Physics and Astronomy, Johns Hopkins University, Baltimore, MD 21218, USA*

[*]Corresponding author: Aghasem2@jhu.edu

We have investigated the effect of synthesis and growth conditions on the magnetic, structural, and compositional properties of pyrochlore oxide holmium titanate and demonstrate a method for growing high quality stoichiometric single crystals. A series of polycrystalline samples with various contents of Ti (-0.08 ≤ x ≤ 0.08, and nominal compositions of Ho$_2$Ti$_{2+x}$O$_7$) were synthesized at different temperatures, and characterized using powder X-ray diffraction. The results show that synthesizing powders at a higher temperature of 1500 °C yield single phase compounds. Ti deficient powders showed an increase of lattice constant due to stuffing (Ho into Ti positions), while Ti rich powders showed a decrease in lattice constant due to anti-stuffing (Ti into Ho positions). A post-annealing in O$_2$ was found to be necessary to accomplish the anti-stuffing process. Use of the conventional floating zone (FZ) technique introduced Ti deficiency, stuffing, and oxygen vacancies in the grown crystal. Growth of high structural quality and stoichiometric single crystals of Ho$_2$Ti$_2$O$_7$ by the traveling solvent floating zone (TSFZ) is reported. AC susceptibility measurements revealed that the stoichiometric crystal shows a higher ice freezing temperature, indicating that crystal quality and stoichiometry play a key role on low temperature spin ice properties of this compound.

## I. INTRODUCTION

The pyrochlore rare earth titanates ($RE_2$Ti$_2$O$_7$, space group $Fd\bar{3}m$) has been the subject of extensive studies over the past two decades [1–11]. In the field of magnetism, this class of materials is considered as the ideal hosts for archetypal geometrical frustration which consists of the quantum spin ice, classical spin ice, and spin liquid behaviors [1,2,4]. These materials have also been studied for their potential use as solid catalysts [5,6], electrolytes [7], ferroelectric, dielectric materials and their magnetic behavior at low temperatures [2]. Although the rare earth titanates have been extensively examined, sample quality varies from study to study, and the impact of structural disorder on the magnetism is poorly understood for many of these compounds.

The pyrochlore structure, A$_2$B$_2$O$_7$, as described by two interpenetrating networks of tetrahedra formed by A and B cations, has been defined as an anion deficient derivative of fluorite structure with two types of ordered cations [9]. The A$^{3+}$ and B$^{4+}$ cations occupy the 16d and 16c sites respectively, and there are two inequivalent oxygen positions, the 48f (where the oxygen atoms are surrounded by two A and two B cations) and the 8b (where the oxygen is surrounded by four A cations). In this structure, the 8a anion site (at the centroids of tetrahedra of 16c B sites) is an oxygen vacancy surrounded by four $B^{4+}$, where the B cations tend to be electrostatically repulsed, and this is compensated for by the displacement of the 48f anions towards the exposed B-site cations [9,12]. The distortion degree of Ti-O octahedron is defined by the *a* axis position of the 48f oxygen, which can be between 0.3125 (where the B ions are in regular octahedral coordination and A ions are in a distorted cubic coordination), and 0.375 (where the coordination of A ions is a perfect cube and Ti octahedron is flattened).

In the case of $Ho_2Ti_2O_7$, $Ho^{3+}$ ions with $4f^{10}$ electron configuration are the source of magnetic behavior, and $Ti^{4+}$ has no valence electrons ($3d^0$) and is thus nonmagnetic in an ideally stoichiometric compound. Due to structural flexibilities of the pyrochlore, Ti atoms can be substituted by other ions such as Zirconium [13,14], Germanium [15–17], Tin [18–20], or even Platinium [21] without altering the crystal structure. Pyrochlores can tolerate vacancies and disorders at A, B and O sites to a certain extent; however, such deviations from the stoichiometric composition have been shown to be responsible for variations of properties [22–25]. Diffuse neutron scattering and magnetization measurements show that oxygen deficiencies affect the low temperature monopole physics of FZ grown crystals [26]. The introduction of oxygen vacancies (mostly from the 48f sites), change of Ti oxidation state to 3+, generation of associated defect structures (e.g. antisite disorders) and formation of two new defect anion positions are discussed in detail in ref. [27]. In addition to point defects, high degrees of atomic disordering, as extended defects of superdislocations and anti-phase boundaries, have been observed in the FZ grown $Yb_2Ti_2O_7$ crystals using the atomic resolution scanning transmission electron microscopy techniques [28].

Holmium titanate samples have been previously prepared in the form of powders [29], small mm size single crystals by the flux method using $PbF_2$ [30], cm size crystals by Czochralski method [31], thin films by pulsed laser deposition (PLD) [32], and larger cm size crystals using the crucible-less floating zone (FZ) technique [33]. Regardless of the preparation technique, however, there are inconsistencies in the reported lattice constants, ranging from 10.009 Å to 10.132 Å [25,30–37]. At a minimum, this raises doubts concerning the stoichiometry and quality of these samples. This is important because off-stoichiometry has shown to significantly affect the magnetic interactions in other pyrochlore titanates [10].

Through a systematic study, we have prepared and characterized a series of polycrystalline samples (with different contents of Ti) to show the effects that preparation conditions have upon the quality, structure and phase purity of $Ho_2Ti_2O_7$. Using powder X-ray diffraction, this study shows that the lattice constant can be used as a measure of the stoichiometry of the sample. We also grew and characterized float-zone grown single crystals, and report a method for growing high quality stoichiometric $Ho_2Ti_2O_7$ crystals by the traveling solvent floating zone technique (TSFZ). Finally, we present evidence from AC susceptibility that the spin ice freezes at higher temperatures in the purest samples.

## II. EXPERIMENTAL PROCEDURE
### A. Synthesis

Powders with nominal compositions of $Ho_2Ti_{2+x}O_7$ (-0.08≤x≤0.08) were synthesized from the starting materials of $Ho_2O_3$ (99.995%, Alpha Aesar) and rutile $TiO_2$ (99.99%, Alpha Aesar) in large batches of 30 grams. The precursors were dried at 900 °C for 10h to minimize the mass error. Powders were mixed and thoroughly ground together in a porcelain mortar and pestle, and then heated in air using alumina crucibles at different temperatures of 1250 °C, 1300 °C, and several times at 1350 °C for 10 h, with an intermediate grinding. These powders were then heated three times more at 1500 °C until powder X-ray diffraction (XRD) showed the precursors were fully reacted.

### B. Single crystal growth

Feed rods of stoichiometric composition were used for growth of single crystals. These polycrystalline feed rods were made by sealing powders in a rubber tube, which were then pressed into rods of about 60 mm in diameter and 70-80 mm in length and sintered at 1350 °C for 10 h in air. Single crystals were grown by both the FZ and TSFZ techniques in a four-mirror halogen optical floating zone furnace (Crystal Systems Inc. FZ-T-4000-H-VII-VPO-PC) with four halogen lamps of 1kW each [38,39]. In all the growths, the seed rod was mounted on the bottom shaft of the furnace and the molten zone moved upward.

### C. Characterization

Powder X-ray diffraction (XRD) data were taken using a Bruker D8 Focus X-ray diffractometer operating with Cu Kα radiation and a LynxEye strip detector. TOPAS software (Bruker AXS) was used to refine and analyze the collected XRD patterns. All the data were taken in the range 5-120° and silicon was added to the samples as a standard of known lattice parameter. Error bars reported for lattice parameters indicate statistical errors. AC susceptibility measurements were performed using a Quantum Design Physical Properties Measurement System (PPMS), with the dilution refrigerator ACMS option. Two cubic single crystals were used for these measurements, a 43 mg crystal cube ($\approx 0.9 \times 0.9 \times 0.9$ mm$^3$) and a 64 mg one ($\approx 1 \times 1 \times 1$ mm$^3$) grown by the FZ and the TSFZ techniques respectively. These samples were mounted on a sapphire rod using GE varnish, and [110] zero-field AC susceptibilities were measured at 10 Hz down to 0.6 K. This data have been corrected for demagnetization factor in accord with reference [40].

### III. RESULTS AND DISCUSSION

### A. Characterization of synthesized powders

XRD patterns of the powder samples ($Ho_2Ti_{2+x}O_7$, x = -0.08, -0.04, 0, 0.04, 0.08), synthesized in air at 1350 °C and 1500 °C, were analyzed via Rietveld refinements using a cubic pyrochlore model to extract lattice parameters, site mixing, and number of phases, as reported in Tables I (a) and (b). Powder X-ray diffractions confirmed the cubic pyrochlore structure for all the samples prepared.

In the powder samples synthesized at 1350 °C, we detected second phases of $TiO_2$ and $Ho_2TiO_5$, for the higher and lower targeted levels of doping, respectively. The existence of these second phases at temperatures below 1400°C has also been reported in the phase diagrams of $Ho_2O_3$-$TiO_2$ [41–43]. The presence of $TiO_2$ second phase and the measured lattice parameter for x≥0 indicate that anti-stuffing (replacement of $Ho^{3+}$ by smaller $Ti^{4+}$ cations) has not occurred at this temperature. In contrast, for lower Ti content powders (x<0), larger lattice parameters were measured, showing that, together with having the second phase $Ho_2TiO_5$, some amount of stuffing (replacement of $Ti^{4+}$ by larger $Ho^{3+}$ cations) has occurred. In the Ti deficient stuffed compounds, $Ho_2(Ti_{2-x}Ho_x)O_{7-\delta}$, oxygen deficient concentration must maintain electrical neutrality of the compound, and this seems to have been corrected in air. Oxygen deficiency and stuffing appear to be interrelated in the pyrochlore structure because the introduction of anion vacancies can also cause antisite disorders [27].

Synthesis of powders at a higher temperature of 1500 °C in air led to the formation of single phase compounds with no evidence of secondary phases, which is consistent with a higher solid solubility at temperatures close to 1500 °C and above, as predicted in the phase diagrams of $Ho_2O_3$-$TiO_2$ [42]. This is confirmed by the observation that the measured lattice parameters of the samples vary systematically with composition in the range of $-0.08 \leq x \leq 0.08$, as reported in Table I(b). A change of lattice parameters was measured from a smaller value of $a = 10.09212(7)$ Å for the Ti rich compound (x = +0.08) as an indication of anti-stuffing, to a larger value of $a = 10.11098(4)$ Å for the highly Ti deficient compound (x = -0.08) due to stuffing. In case of the compound x=+0.04, although no indication of second phases was initially observed for the powder synthesized at 1350 °C, the change of the lattice parameter to a smaller value (following a higher synthesis temperature of 1500 °C) indicates that it should have contained a very small amount of second phase $TiO_2$ which was below the detection limit of our XRD instrument.

The room temperature lattice parameter of $a = 10.09832(1)$ Å corresponds to the stoichiometric $Ho_2Ti_2O_7$ reference powder, as XRD pattern is shown in Fig. 1. A lattice constant of 10.1 Å [32] reported for stoichiometric holmium titanate seems to be close to our reference powder, while considerable variability from small values of 10.07 [30] to large values of 10.132 [37] can be seen in other studies.

In contrast to studies on stuffing [5,44–49], anti-stuffing or negative stuffing has been rarely discussed and agreed on in the literature [50]. In our study, for powders of higher levels of Ti doping (x>0) synthesized at 1500 °C in air, change of lattice parameters to smaller values was observed as an indication of anti-stuffing. Further analysis of XRD pattern for x = + 0.08 showed peak splitting which can be attributed to two similar pyrochlore phases with very close lattice parameters (Fig. 2). Therefore, the powder sample was additionally annealed at 1350 °C in $O_2$, and this restored the peak shape, as shown in Fig. 2. This is consistent with the charge neutrality of the anti-stuffed compounds, $(Ho_{2-x}Ti_x)Ti_2O_{7+\delta}$, since they need an oxygen rich atmosphere to be developed, as opposed to the stuffed compounds, $Ho_2(Ti_{2-x}Ho_x)O_{7-\delta}$, for which oxygen deficient environments of air and argon are more favorable. The XRD refined information for the anti-stuffed powder (x = 0.08) is given in Table II where annealing in oxygen has corrected the lattice constant. For the powder with a rather lesser doping level of Ti (x = 0.04), a noticeable correction of the lattice constant after oxygen annealing was not observed within the detection limit of our X-ray diffraction instrument, and Ho site occupancies were found to be nearly one.

### B. Crystal growth by float-zone melting techniques

Crystal growth was initially attempted by the conventional floating zone (FZ) technique, and a crystal of dark-orange color, as not expected for $Ho^{3+}$ and $Ti^{4+}$, was grown (Fig. 3). Since Ti can adapt two different valance states ($Ti^{4+}$ and $Ti^{3+}$), the dark color is attributed to the intervalence charge transition to $Ti^{3+}$ due to oxygen vacancies as mainly caused by growth at high temperatures at and above the melting point of $Ho_2Ti_2O_7$ [40,49,51,52]. Very small amount of $Ti^{3+}$ ions can noticeably change the color of a clear/transparent material of single crystal form which can be less evident in the powder form [53]. The $Ti^{3+}$ concentration, which depends on the growth conditions (e.g. temperature, atmosphere, zoning rate), can be different in the FZ grown crystals; therefore, colors can vary from a rich orange to a dark color. In general, the fact that the crystal changes color, which can be explained in terms of crystal-field splitting of the energy levels of the $Ti^{3+}$ ion, indicates the existence of magnetic $Ti^{3+}$ and oxygen vacancies in the structure [26,41].

We observed evidence of titanium oxides second phases, as seen for $Yb_2Ti_2O_7$ [10], in the XRD analysis of the molten zone area, confirming that the $Ho_2Ti_2O_7$ single crystal is not a congruently melting compound. This was not predicted in the reported phase diagrams [41–43]. In this regard, the cubic lattice parameter revealed a large gradient along the grown crystal, ranging from 10.10384(7) Å at the start of the grown crystal (cut 1) consistent with the Ti deficient or Ho stuffed powders, to 10.09951(9) Å at the end of the crystal (cut 3) which is close to the stoichiometric powder. Refined X-ray diffraction data, including the lattice constants and systematic change of Ho occupancy in Ti sites as a sign of stuffing in different sections of the FZ grown crystals, are reported in Table III.

Post-growth annealing of dark color FZ crystals in oxygen can correct the oxidation state of Ti (Fig. 3); however, this process seems to have an insignificant effect on the other existing defects (e.g. stuffing, and disorders) as only small changes were seen in the lattice parameter and occupancies (of Ti and Ho), shown in Table III.

Despite the marked variations in color and lattice constants, however, X-ray Laue pictures taken at regular intervals along the lengths of the crystal indicated that it exhibited an equally high crystalline quality with no detectable variation of the orientation between pictures and no evidence of spot splitting or distortion (the X-ray beam used had a diameter of 1mm and allowed orientation variations of less than 1° to be detected).

Based on the eutectic line shown on the $Ho_2O_3$-$TiO_2$ phase diagram [41–43], the traveling solvent floating zone (TSFZ) technique, a powerful growth technique for incongruently melting compounds [38,39], was used to grow crystals of $Ho_2Ti_2O_7$. This was accomplished using a $TiO_2$ rich solvent at a temperature lower than the melting point of $Ho_2Ti_2O_7$. A 5 cm long transparent orange color single crystal grown using this technique is pictured in Fig. 4. The lattice constant along the crystal was nearly identical and measured to be $a$ = 10.09890(7) Å, reported in Table III, and this matches well the lattice constant of the stoichiometric powder. An equally high crystalline quality with no detectable variation of the orientation was seen at the X-ray Laue pictures taken at regular intervals along the length of the crystal.

Lattice constant measurements of stuffed (x<0), stoichiometric, and anti-stuffed powders (x>0) are shown in Fig. 5 together with those of crystals grown by FZ and TSFZ techniques. Based on the larger lattice constants measured at different parts of the FZ grown crystals, Ti deficiency and stuffing of different levels is considered to be present in these crystals, in addition to oxygen deficiencies and $Ti^{3+}$ ions as discussed earlier. More importantly, lattice constants of the reference stoichiometric powder and the TSFZ grown crystal are equivalent, indicating the high quality and stoichiometry of the prepared crystal using this method. To the best of our knowledge this is the first stoichiometric crystal grown by the TSFZ technique.

### C. Susceptibility Measurements

Susceptibility measurements show that the sample quality of holmium titanate has an important effect on the low temperature physical properties, particularly when it enters the spin ice phase. Theoretical models have shown that defect levels of 0.3% affect the transport properties of magnetic monopoles within the spin-ice phase, suggesting that reducing defect levels "could alter the transport properties in fundamental ways" [54]. Indications of this difference in the AC susceptibility data can be seen in the Figure 5. $\chi$" peaks at the highest temperatures when the samples are the closest to stoichiometric, suggesting that stuffing disorder suppresses the freezing into the spin-ice phase. This is consistent with the explanation offered in ref. [54] that the monopoles get "trapped" in the vicinity of site defects, such that removing the defects allows for freezing at higher temperatures. This result also suggests that disagreements in the low-temperature susceptibility [40] may be due to sample quality, and not just sample orientation or measurement technique. This result has important implications for quantitative comparisons between experiment and theory. In as much as it has been a challenge to characterize the relaxation and transport properties of spin ice compounds [40,54–58], it is vitally important that the effects of defects not be confused with intrinsic spin-ice behaviors. For any quantitative measurements of monopole activation energy and dynamics, it will be necessary to carefully account for sample quality.

### IV. CONCLUSIONS

By preparing a series of holmium titanate powders at different Ti levels, we have shown that the structural quality of $Ho_2Ti_2O_7$ is very sensitive to synthesis conditions and small deviations from stoichiometry, as these can lead to lattice constant variations, site disorders, vacancy defects, and second phase inclusions. Larger lattice constants were measured in Ti deficient powders as an indication of stuffing, while in anti-stuffed Ti-rich powders the lattice constants were smaller. Single crystals grown by the floating zone (FZ) was shown to have stuffing of various amounts (depending on the Ti deficiency level along the crystal), and oxygen deficiencies as it was dark in color due to $Ti^{3+}$ inclusions. Using $TiO_2$ as the solvent, the travelling solvent floating zone technique (TSFZ) was successfully carried out and a high structural quality, well stoichiometric, and transparent orange color single crystal of $Ho_2Ti_2O_7$ was grown for the first time. Low temperature AC susceptibility measurements revealed that the stoichiometric TSFZ crystal freezes into the ice phase at a higher temperature than the less pure FZ crystal, indicating that the disorder and defects work against the intrinsic spin-ice physics, and any quantitative comparison between measurement and theory must take this into account. This result highlights the importance of having high quality crystals when probing the behavior of classical spin ice.

### V. ACKNOWLEDGMENT

This work was supported by U.S. Department of Energy (DOE), Office of Basic Energy Sciences, Division of Materials Sciences and Engineering under award DE-FG02-08ER46544. A.S. and J.K. were supported through the Gordon and Betty Moore foundation under the EPIQS program GBMF4532. The author would like to acknowledge Benjamin A. Trump for helpful discussions.**References:**
[1]   A. P. Ramirez, Annu. Rev. Mater. Sci. **24**, 453 (1994).
[2]   J. S. Gardner, M. J. P. Gingras, and J. E. Greedan, Rev. Mod. Phys. **82**, 53 (2010).
[3]   A. Scheie, J. Kindervater, S. Säubert, C. Duvinage, C. Pfleiderer, H. J. Changlani, S. Zhang, L. Harriger, K. Arpino, S. M. Koohpayeh, O. Tchernyshyov, and C. Broholm, Phys. Rev. Lett. **119**, 1 (2017).
[4]   S. T. Bramwell and M. J. P. Gingras, Science (80-. ). **294**, 1495 (2001).
[5]   J. Lian, J. Chen, M. Wang, R. C. Ewing, J. M. Farmer, L. A. Boatner, and B. Helean, Phys. Rev. B - Condens. Matter Mater. Phys. **68**, 1 (2003).
[6]   S. H. Oh, R. Black, E. Pomerantseva, J. H. Lee, and L. F. Nazar, Nat. Chem. **4**, 1004 (2012).
[7]   A. V Shlyakhtina and L. G. Shcherbakova, Russ. J. Electrochem. **48**, 1 (2012).
[8]   S. Rosenkranz, A. P. Ramirez, A. Hayashi, R. J. Cava, R. Siddharthan, and B. S. Shastry, J. Appl. Phys. **87**, 5914 (2000).
[9]   M. A. Subramanian, G. Aravamudan, and G. V. S. Rao, Rev. Lit. Arts Am. **15**, 55 (1983).
[10]  K. E. Arpino, B. A. Trump, A. O. Scheie, T. M. McQueen, and S. M. Koohpayeh, Phys. Rev. B **95**, 094407 (2017).
[11]  C. Nisoli, R. Moessner, and P. Schiffer, Rev. Mod. Phys. **85**, 1473 (2013).
[12]  J. M. Farmer, L. A. Boatner, B. C. Chakoumakos, M. H. Du, M. J. Lance, C. J. Rawn, and J. C. Bryan, J. Alloys Compd. **605**, 63 (2014).
[13]  H. Y. Xiao, F. Gao, and W. J. Weber, Phys. Rev. B - Condens. Matter Mater. Phys. **80**, 1 (2009).
[14]  R. Clements, J. R. Hester, B. J. Kennedy, C. D. Ling, and A. P. J. Stampfl, J. Solid State Chem. **184**, 2108 (2011).
[15]  A. M. Hallas, J. A. M. Paddison, H. J. Silverstein, A. L. Goodwin, J. R. Stewart, A. R. Wildes, J. G. Cheng, J. S. Zhou, J. B. Goodenough, E. S. Choi, G. Ehlers, J. S. Gardner, C. R. Wiebe, and H. D. Zhou, Phys. Rev. B - Condens. Matter Mater. Phys. **86**, 1 (2012).
[16]  S. Jana, D. Ghosh, and B. M. Wanklyn, J. Magn. Magn. Mater. **183**, 135 (1998).
[17]  D. M. Moran, F. S. Richardson, M. Koralewski, and B. M. Wanklyn, J. Alloys Compd. **180**, 171 (1992).
[18]  H. Kadowaki, Y. Ishii, K. Matsuhira, and Y. Hinatsu, Phys. Rev. B **65**, 1 (2001).
[19]  K Matsuhira et al, J. Phys. Condens. Matter **12**, L649 (2000).
[20]  G. Ehlers, A. Huq, S. O. Diallo, C. Adriano, K. C. Rule, A. L. Cornelius, P. Fouquet, P. G. Pagliuso, and J. S. Gardner, J. Phys. Condens. Matter **24**, 076005 (2012).
[21]  J. Ostorero, H. Makram, and T. Rares, J. Cryst. Growth **25**, 677 (1974).
[22]  C. R. Stanek, L. Minervini, and R. W. Grimes, J. Am. Ceram. Soc. **85**, 2792 (2002).
[23]  L. Minervini, R. W. Grimes, Y. Tabira, R. L. Withers, and K. E. Sickafus, Philos. Mag. A Phys. Condens. Matter, Struct. Defects Mech. Prop. **82**, 123 (2002).
[24]  Y. Tabira, R. L. Withers, L. Minervini, and R. W. Grimes, J. Solid State Chem. **153**, 16 (2000).
[25]  G. C. Lau, R. S. Freitas, B. G. Ueland, B. D. Muegge, E. L. Duncan, P. Schiffer, and R. J. Cava, Nat. Phys. **2**, 249 (2006).
[26]  G. Sala, M. J. Gutmann, D. Prabhakaran, D. Pomaranski, C. Mitchelitis, J. B. Kycia, D. G. Porter, C. Castelnovo, and J. P. Goff, Nat. Mater. **13**, 488 (2014).
[27]  G. D. Blundred, C. A. Bridges, and M. J. Rosseinsky, Angew. Chemie - Int. Ed. **43**, 3562 (2004).
[28]  Z. Shafiezadeh, Y. Xin, Q. Huang, H. Zhou, and S. M. Koohpayeh, Submitted Nat. Phys. (2018).
[29]  L. H. Brixner, Inorg. Chem. **3**, 1065 (1964).
[30]  J. D. Cashion,  a. H. Cooke, M. J. M. Leask, T. L. Thorp, and M. R. Wells, J. Mater. Sci. **3**, 402 (1968).7

[31] J. Kang, W. Xu, W. Zhang, X. Chen, W. Liu, F. Guo, S. Wu, and J. Chen, J. Cryst. Growth **395**, 104 (2014).
[32] D. P. Leusink, F. Coneri, M. Hoek, S. Turner, H. Idrissi, G. Van Tendeloo, and H. Hilgenkamp, APL Mater. **2**, 032101 (2014).
[33] Q. J. Li, L. M. Xu, C. Fan, F. B. Zhang, Y. Y. Lv, B. Ni, Z. Y. Zhao, and X. F. Sun, J. Cryst. Growth **377**, 96 (2013).
[34] A. L. Cornelius and J. S. Gardner, Phys. Rev. B - Condens. Matter Mater. Phys. **64**, 604061 (2001).
[35] D. Prabhakaran and A. T. Boothroyd, J. Cryst. Growth **318**, 1053 (2011).
[36] K. Baroudi, B. D. Gaulin, S. H. Lapidus, J. Gaudet, and R. J. Cava, Phys. Rev. B **92**, 024110 (2015).
[37] C. Krey, S. Legl, S. R. Dunsiger, M. Meven, J. S. Gardner, J. M. Roper, and C. Pfleiderer, Phys. Rev. Lett. **108**, 1 (2012).
[38] S. M. Koohpayeh, D. Fort, and J. S. Abell, Prog. Cryst. Growth Charact. Mater. **54**, 121 (2008).
[39] S. M. Koohpayeh, Prog. Cryst. Growth Charact. Mater. **62**, 22 (2016).
[40] J. A. Quilliam, L. R. Yaraskavitch, H. A. Dabkowska, B. D. Gaulin, and J. B. Kycia, Phys. Rev. B - Condens. Matter Mater. Phys. **83**, 28 (2011).
[41] G. V. Shamrai, R. L. Magunov, L. V. Sadkovskaya, I. V. Stasenko, and I. P. Kovalevskaya, Inorg. Mater. (Translated from Neorg. Mater. **27**, 140 (1991).
[42] G. E. Sukhanova, K. N. Guseva, A. V Kolesnikov, and L. G. Shcherbakova, Inorg. Mater. (Engl. Transl.) **18**, 2014 (1982).
[43] Z. K. Huang, Z. X. Lin, and T. S. Yen, Guisuanyan Xuebao **7**, 1 (1979).
[44] Y. H. Li, B. P. Uberuaga, C. Jiang, S. Choudhury, J. A. Valdez, M. K. Patel, J. Won, Y. Q. Wang, M. Tang, D. J. Safarik, D. D. Byler, K. J. McClellan, I. O. Usov, T. Hartmann, G. Baldinozzi, and K. E. Sickafus, Phys. Rev. Lett. **108**, 1 (2012).
[45] J. Chen, J. Lian, L. M. Wang, R. C. Ewing, R. G. Wang, and W. Pan, Phys. Rev. Lett. **88**, 105901 (2002).
[46] J. Lian, L. Wang, J. Chen, K. Sun, R. C. Ewing, J. M. Farmer, and L. A. Boatner, Acta Mater. **51**, 1493 (2003).
[47] P. Telang, K. Mishra, A. K. Sood, and S. Singh, arXiv: 1805.01674, (2018).
[48] K. Nakamura, M. Mori, T. Itoh, and T. Ohnuma, AIP Adv. **6**, 115003 (2016).
[49] D. J. Gregg, Z. Zhang, G. J. Thorogood, B. J. Kennedy, J. A. Kimpton, G. J. Griffiths, P. R. Guagliardo, G. R. Lumpkin, and E. R. Vance, J. Nucl. Mater. **452**, 474 (2014).
[50] A. Mostaed, G. Balakrishnan, M. R. Lees, and R. Beanland, Acta Mater. **143**, 291 (2018).
[51] S. M. Koohpayeh, D. Fort, and J. S. Abell, J. Cryst. Growth **282**, 190 (2005).
[52] X. Chen, L. Liu, and F. Huang, Chem. Soc. Rev. **44**, 1861 (2015).
[53] R. J. D. Tilley, *Colour and the Optical Properties of Materials*, Second (John Wiley & Sons, Ltd, Chichester, UK, 2010).
[54] H. M. Revell, L. R. Yaraskavitch, J. D. Mason, K. A. Ross, H. M. L. Noad, H. A. Dabkowska, B. D. Gaulin, P. Henelius, and J. B. Kycia, Nat. Phys. **9**, 34 (2013).
[55] A. B. Eyvazov, R. Dusad, T. J. S. Munsie, H. A. Dabkowska, G. M. Luke, E. R. Kassner, J. C. S. Davis, and A. Eyal, arXiv: 1707.09014 (2017).
[56] D. Pomaranski, L. R. Yaraskavitch, S. Meng, K. A. Ross, H. M. L. Noad, H. A. Dabkowska, B. D. Gaulin, and J. B. Kycia, Nat. Phys. **9**, 353 (2013).
[57] S. R. Giblin, M. Twengström, L. Bovo, M. Ruminy, M. Bartkowiak, P. Manuel, J. C. Andresen, D. Prabhakaran, G. Balakrishnan, E. Pomjakushina, C. Paulsen, E. Lhotel, L. Keller, M. Frontzek, S. C. Capelli, O. Zaharko, P. A. McClarty, S. T. Bramwell, P. Henelius, and T. Fennell, arXiv: 1804.08970 (2018).
[58] L. D. C. Jaubert and P. C. W. Holdsworth, J. Phys. Condens. Matter **23**, (2011).

TABLE I. Refined lattice constants, observed phases, O(48f) positional parameters, and occupancy sites of the powder samples with the nominal compositions of $Ho_2Ti_{2+x}O_7$, synthesized at 1350°C, and additionally at 1500°C in air.

(a) Synthesized at 1350 °C

| x | Lattice Constant (Å) | Observed Phases Main Phase | Observed Phases Second Phases | O(48f) x | Occupancy of Ti in Ti Site | Occupancy of Ho in Ti Site | $x^2$ |
|---|---|---|---|---|---|---|---|
| 0.08 | 10.09949(6) | $Ho_2Ti_2O_7$ | $TiO_2$ | 0.31636 | 1 | 0 | 2.71(3) |
| 0.04 | 10.09958(6) | $Ho_2Ti_2O_7$ | - | 0.31508 | 1 | 0 | 5.11(1) |
| 0 | 10.09993(0) | $Ho_2Ti_2O_7$ | - | 0.31996 | 1 | 0 | 2.44(6) |
| -0.04 | 10.10183(9) | $Ho2(Ti_{2-x}Ho_x)O_{7-\delta}$ | $Ho_2TiO_5$ | 0.31721 | 0.9404 | 0.0596 | 3.98(3) |
| -0.08 | 10.10364(9) | $Ho2(Ti_{2-x}Ho_x)O_{7-\delta}$ | $Ho_2TiO_5$ | 0.31720 | 0.9248 | 0.0752 | 3.22(2) |

(b) Synthesized at 1500 °C

| x | Lattice Constant (Å) | Observed Phases Main Phase | Observed Phases Second Phases | O(48f) x | Occupancy of Ti in Ti Site | Occupancy of Ho in Ti Site | $x^2$ |
|---|---|---|---|---|---|---|---|
| 0.08 | 10.09212(7) | $(Ho_{2-x}Ti_x)Ti_2 O_{7+\delta}$ | - | 0.31392 | 1 | 0 | 3.60(0) |
| 0.04 | 10.09758(2) | $(Ho_{2-x}Ti_x)Ti_2 O_{7+\delta}$ | - | 0.31297 | 1 | 0 | 3.73(3) |
| 0 | 10.09832(1) | $Ho_2Ti_2O_7$ | - | 0.31792 | 1 | 0 | 3.57(8) |
| -0.04 | 10.10421(3) | $Ho2(Ti_{2-x}Ho_x)O_{7-\delta}$ | - | 0.31723 | 0.9330 | 0.0670 | 3.12(7) |
| -0.08 | 10.11098(4) | $Ho2(Ti_{2-x}Ho_x)O_{7-\delta}$ | - | 0.31715 | 0.9217 | 0.0783 | 3.65(2) |

TABLE II. Lattice constant, O(48f) x position, and occupancy of Ho site refined for the Ti-rich (x=0.08) powder samples, as synthesized in air at 1500 °C, and after annealing in oxygen at 1350 °C.

| X = 0.08 | Lattice Constant (Å) | O(48f) x | Occupancy of Ho in Ho Site | Occupancy of Ti in Ho Site | $x^2$ |
|---|---|---|---|---|---|
| As synthesized | 10.09212(7) | 0.31392 | 0.923 | 0.077 | 3.60(0) |
| After annealing in O$_2$ | 10.09364(6) | 0.31279 | 0.922 | 0.078 | 2.97(2) |

TABLE III. Unit cell, O(48f) x position, and occupancy of Ti site as refined for the crystals grown by FZ and TSFZ techniques.

| Sample | Lattice parameter(Å) | O (48f) x | Occupancy of Ti in Ti Site | Occupancy of Ho in Ti Site | $x^2$ | Number of Phases |
|---|---|---|---|---|---|---|
| TSFZ | 10.09890(7) | 0.32048 | 1 | 0 | 3.490 | 1 |
| FZ-Cut 1 | 10.10384(7) | 0.32229 | 0.972 | 0.028 | 1.524 | 1 |
| FZ-Cut 2 | 10.10135(9) | 0.32114 | 0.989 | 0.011 | 1.898 | 1 |
| FZ-Cut 3 | 10.09951(9) | 0.32164 | 1 | 0 | 1.674 | 1 |
| FZ-Cut 1 Annealed | 10.10396(6) | 0.31709 | 0.9825 | 0.0175 | 2.168 | 1 |

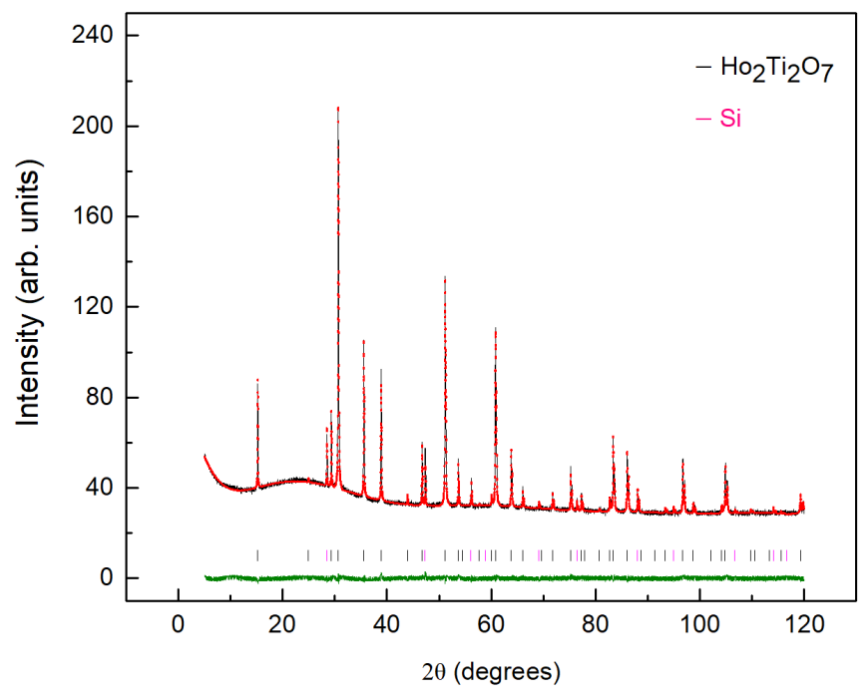

FIG. 1. X-ray diffraction pattern of the stoichiometric $Ho_2Ti_2O_7$ powder sample confirming the pyrochlore structure, with lattice constant of a = 10.09832(1) Å. Black line is observed intensity, the red circles are calculated intensity, and the green line is the difference. Si was added as the standard with hkl reflections in pink.

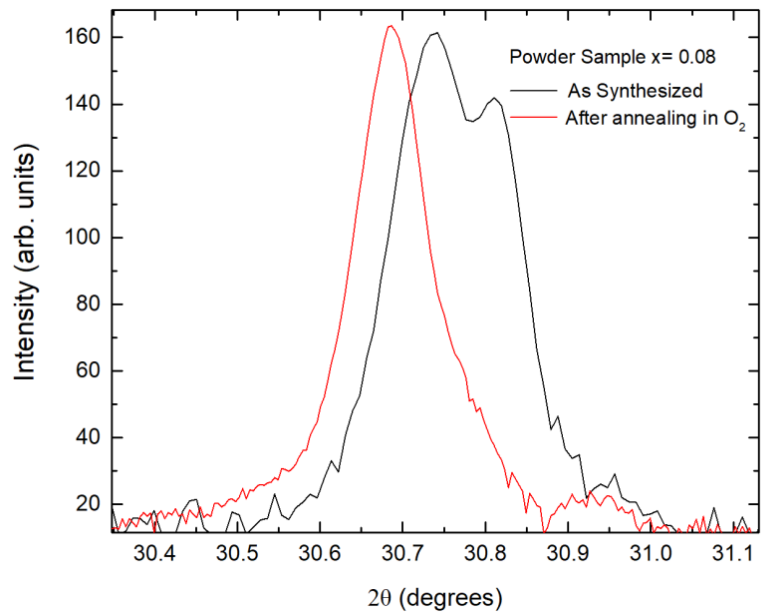

FIG. 2. Powder X-ray diffraction patterns, shown in the 2θ range of about 30-31 degrees, for the powder sample of x = 0.08 as synthesized in air (black), and after annealing in $O_2$ (red). The peak shape is restored due to annealing in oxygen, confirming the formation of a single pyrochlore structure.

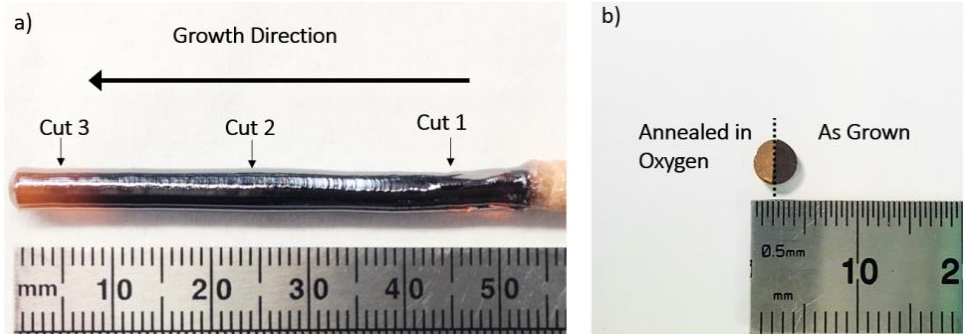

FIG.3. (a) Dark-color stuffed single crystal grown by conventional floating zone technique. Lattice constants along the growth direction were shown a reduction from the start of growth (cut 1) to the end (cut 3). (b) A disk, from the beginning of growth, was cut into half, and one piece was annealed in $O_2$ at 1000 °C. This corrected the color, from dark to orange color, while the lattice constant did not change considerably.

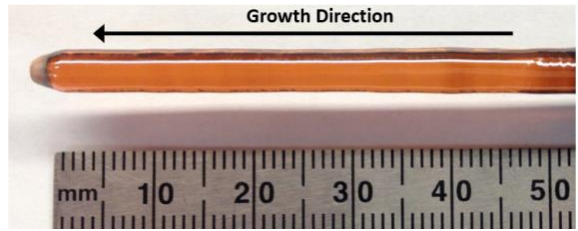

FIG. 4. Stoichiometric single crystal of $Ho_2Ti_2O_7$, with a consistent lattice constant of a = 10.09890(7) Å, was grown by traveling solvent floating zone (TSFZ) technique using an image furnace.

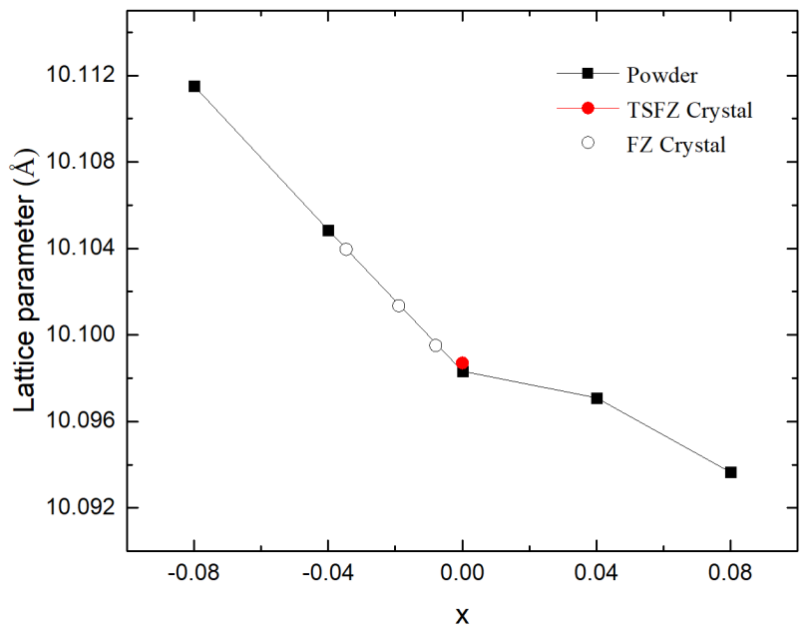

FIG. 5. Composition dependence of the lattice constants for different Ti doping levels, with nominal compositions Ho$_2$Ti$_{2+x}$O$_7$. Measured lattice constants for the stuffed Ti deficient were larger, while for the anti-stuffed Ti rich samples smaller values were obtained. FZ grown crystals showed Ti deficiencies of different concentrations along the crystal, with the largest Ti deficiency at the start of crystal, and lower deficiencies for the end part of crystal. Lattice parameter obtained for the high quality and stoichiometric TSFZ grown crystal matches the stoichiometric reference powder.

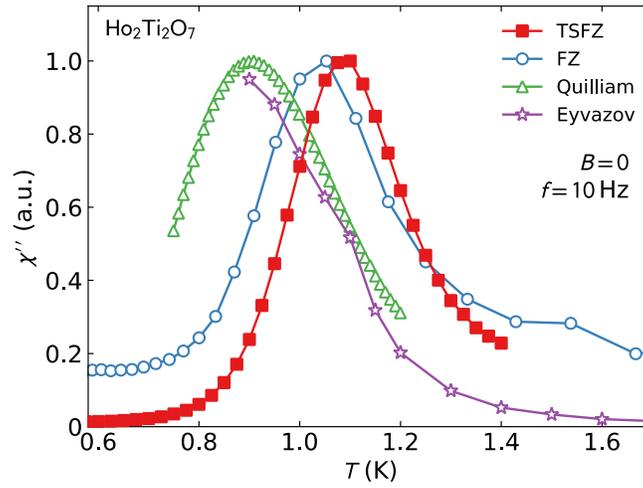

FIG.6. Imaginary component of AC susceptibility at 10 Hz as a function of temperature for $Ho_2Ti_2O_7$. The peak indicates the onset of spin-ice freezing. The stoichiometric TSFZ sample χ" peak is at 1.09 K, while the FZ sample χ" peak is at 1.04 K. Also plotted is data from ref. [40] and [55] ,showing peaks at 0.91 K and < 0.90 K, respectively.